\documentstyle[preprint,aps]{revtex}
\begin{document}
\draft
\title{Escape from Metastability via Aging:
Non-Equilibrium Dynamics in a One-Dimensional Ising Model}
\author{J.\ Kisker, H.\ Rieger and M.\ Schreckenberg}
\address{
Institut f\"ur Theoretische Physik\\
Universit\"at zu K\"oln\\
50937 K\"oln,Germany
}
\date{\today}
\maketitle
\begin{abstract}
The non-equilibrium dynamics of a one-dimensional Ising model 
with uniform, short-ranged three-spin interactions is investigated.
It is shown that this model possesses an exponentially large 
number of metastable 
configurations that are stable against single spin flips. This 
glass-like situation results in a complete freezing of the system 
at low temperatures for times smaller than an intrinsic time-scale, 
which diverges exponentially with inverse temperature. Via thermal 
activation the system eventually escapes from this frozen state,
which is signals the onset of aging by domain growth.
\end{abstract}
\pacs{75.10N, 75.50L, 75.40G.}
Glassy dynamics is one of the most fascinating subjects in 
modern physics \cite{glass1,glass2}.
It manifests itself in an extremely slow relaxation
exceeding laboratory time-scales caused by a rough energy landscape.
In spin glasses this complex dynamics is due to frustration and
randomness \cite{sgrev} and results in well known aging effects 
\cite{struik,agerev} meaning a characterstic history dependence of 
dynamical observations.

However, as is obvious from the situation in e.g.\ window glass, disorder
is not a necessary ingredience for these phenomena to occur. For
spin models it has been pointed out long time ago \cite{villain}
that frustration without disorder might be able to produce
a low temperature behavior reminiscent of spin glasses. Indeed,
experiments with geometrically frustrated antiferromagnets \cite{exp1,exp2}
showed anomalous dynamical behavior below a certain temperature
that was interpreted as a spin glass transition. The magnetic properties
of these materials are supposed to be adequately described by 
antiferromagnetically-coupled Heisenberg spins on a Kagom\'e lattice and 
recent numerical studies of this model also yield indications for dynamical
freezing \cite{reimers} and possibly spin glass ordering \cite{chandra}
at low temperatures.

On the other side, also disorder without frustration can cause
anomalous dynamics and slow relaxation: Diluted or disordered
ferromagnets are prominent examples \cite{oned,chow} for systems
in which non-equilibrium dynamics is characterized by domain
growth that is drastically slowed down by pinning of
domain walls at vacancy-sites or weak bonds. Obviously at low
enough temperatures it becomes very hard to discriminate such a
scenario from true spin glass dynamics. Even non-frustrated, 
non-disordered models at or close to criticality will exhibit
anomalously slow non-equilibrium relaxation including aging,
originally supposed to be typical for spin glasses 
\cite{oned,cugpure}.

In this short note we present a very simple spin model without
any frustration or disorder that possesses many glass like features
at low temperatures. It turns out that quenching the system from 
high to low temperatures the dynamics gets stucked within {\it microscopic} 
time scales in a metastable configuration without long range order. 
The system remains frozen in this amorphous state for a {\it macroscopic} 
time before it will try to reach an energetically more favorable,
ordered state via domain growth. This is very reminiscent of the
structural glass transition scenario \cite{glass2}, however this
should only be taken as a pictorial analogy.

The model which we consider is a one-dimensional Ising spin system
with $p$-spin interactions defined by the Hamiltonian
\begin{equation}
H=-J\sum_{i=1}^L \,S_i S_{i+1} \cdots S_{i+p-1}
\;,\qquad(S_i=\pm1)\;.
\label{hamil}
\end{equation}
The constant $J$ fixes the energy scale of the spin interactions, 
which we set to one from now on, and
periodic boundary conditions are imposed. The dynamics is defined to
be the usual Glauber-dynamics \cite{glauber}, where each spin is flipped
with probability 
\begin{equation}
w(S_i\to-S_i)= \frac{1}{2}[1-S_i\tanh(h_i/T)]\;,
\label{prob}
\end{equation}
where the local field $h_i$ is defined to be one half of the energy
difference between the configuration with spin $S_i=+1$ and the same
configuration with spin $S_i=-1$, and $T$ is the temperature. We use
sequential update for the analytic calculations as well as for the
numerical simulations.  For $p=2$ one gets the well known
ferromagnetic Ising chain, whose dynamics with random sequential
update has been solved by Glauber \cite{glauber}.  Note that for $p$
odd the Hamiltonian (\ref{hamil}) does not have the usual spin-flip
symmetry $S_i\to-S_i$. Furthermore the local field is a sum over $p$
terms with values $+1$ or $-1$, from which it follows that it can
never be identical to zero in the case $p$ odd, whereas for even $p$
the probability for a spin to have zero local field is finite. Thus
one has to discriminate between two different situations.  For $p$ odd
a new situation arises, similar to $p=2$ with external field
\cite{rem}. For $p$ even and $p\ge4$ features of the latter case will
coexist with those already known from the case $p=2$ \cite{glauber}.
Therefore we will concentrate on the most simple case with $p$ odd,
i.e.\ $p=3$ from now on and leave the study of $p\ge4$ to a more
detailed analysis \cite{big}.

First we study the zero-temperature properties of this model.
The groundstate of this system has a 4--fold degeneracy (in general 
$2^{p-1}$) \cite{remark} given by
\begin{equation}
\begin{array}{cc}
(1) & \cdots+++++++++ \cdots\\
(2) & \cdots+--+--+-- \cdots\\
(3) & \cdots-+--+--+- \cdots\\
(4) & \cdots--+--+--+ \cdots
\end{array}
\end{equation}
Introducing a local energy-variable $\tau_i = S_{i-1}S_i S_{i+1}$
all groundstate configurations are described by $\tau_i=+1$ for all 
sites $i$. Consider a configuration in $\tau$ and $S$ variables
\begin{equation}
\begin{array}{cccc}
(\underline{\tau}) & \cdots +++++++++++&-&+++++++++++++++ \cdots\\
(\underline{S})    & \cdots +++++++++++&+&--+--+--+--+--+ \cdots\\
                  & & i &  
\end{array}
\label{config}
\end{equation}
which consists of two domains, both being in a minimum
energy configuration, separated by a domain wall located at site $i$
(note that in general one has $2^{p-1}$ different kinds of domains, 
corresponding to the ground state degeneracy). A closer look to the
transition probabilities (\ref{prob}) tells us that it
costs an energy amount of $2S_j h_j=2(\tau_{j-1}+\tau_j+\tau_{j+1})$
to flip one spin, i.e.\ in configuration 
(\ref{config}) $6J$ for spins within the domains
and $2J$ to flip the spins at the three sites $i-1$, $i$ and $i+1$ 
surrounding the domain wall. Thus this configuration is stable 
against single spin flips and at finite, but small temperatures 
it needs a time 
\begin{equation}
\tau_{\rm freeze}\approx\exp(2J/T)
\end{equation}
to move the domain wall one lattice spacing to the left or right. 
This is the time scale (which is quite large for $T\ll J$) on which the 
particular configuration remains frozen for small temperatures. Moreover,
it can be shown that all configurations of the type 
(\ref{config}) consisting (expressed in $\tau$-variables) of
strings of arbitrary length $l\ge2$ with $\tau=+1$ separated by isolated sites
with $\tau=-1$ are indeed metastable: From what has been said above it is
clear that the criterion for metastability is 
$\tau_{i-1}+\tau_i+\tau_{i+1}>0$ for all sites $i$. Thus in any triplet
$(\tau_{i-1},\tau_i,\tau_{i+1})$ at most one minus sign may occur.
Hence any metastable configuration can be represented by an arbitray sequence
of two elementary units "$+-+$" and "$+$". The total number $n_L$ 
of all possible sequences in a system with $L$ sites can be calculated via
the Fibonacci-like iteration $n_L=n_{L-1}+n_{L-3}$. The result for
$L\to\infty$ is
\begin{equation}
n_L\sim x^L\quad{\rm with}\quad x=\sqrt[3]{\frac{29+\sqrt{837}}{54}}
+\sqrt[3]{\frac{29-\sqrt{837}}{54}}\approx1.46557\;.
\end{equation}
Starting with a random
initial state the sequential update procedure at zero temperature
will drive the system into one of this exponentially large number
of metastable states within only two sweeps through the
whole chain. A random sequential update and small nonvanishing temperatures
will not change this scenario significantly. Thus after $2\tau_0$, where
$\tau_0$ is the microscopic time scale, the system will be frozen for
a time $\tau_{\rm freeze}=\tau_0\exp(2J/T)$.

What has been said so far can be quantified by looking at the spin
autocorrelation function $C_T(t,t_w)=1/L\sum_{i=1}^L \langle
S_i(t+t_w) S_i(t_w)\rangle_T$ and the time-dependent energy
$E_T(t)=1/L\sum_{i=1}^L \langle \tau_i(t) \rangle_T$ where
$\langle\cdots\rangle_T$ means the expectation value with respect to
the (time-dependent) probability distribution of spin configurations
determined by the Master equation for the stochastic process
considered here (in the limit $L\to\infty$).  These quantities can
easily be calculated for the ferromagnetic Ising chain (i.e.\ $p=2$)
\cite{glauber}).  In the present case such a treatment is not possible
for the same reasons as in the ferromagnetic Ising chain {\it in an
external field} or the Cayley tree with branching number two
\cite{rem}.  However, the remanent magnetization $C_T(t,t_w)$ and the
energy $E_T(t)$ can be calculated analytically for zero temperature
with the tools introduced in \cite{rem}.  In this letter we only note
that an important ingredience for this problem to be exactly solvable
is the fact that the local field acting on the spins never vanishes
(details will be published elsewhere \cite{big}):
\begin{equation}
\begin{array}{ccll}
C_{T=0}(t,0) & = & 0.475 & {\rm for}\quad t\ge2\\
C_{T=0}(1,0) & = & 0.5 & \\
C_{T=0}(1,1) & = & 0.9 & \\
C_{T=0}(t,t_w) & = & 1 &{\rm for}\quad t_w\ge2
\end{array}
\quad
\begin{array}{ccll}
E_{T=0}(1) & = & -0.5 & \\
E_{T=0}(t) & = & -0.6 &{\rm for}\quad t\ge2
\end{array}
\label{zero}
\end{equation}

According to the above arguments the relations (\ref{zero})
also hold for $T\ne0$ as long as $t\ll\tau_{\rm freeze}$. 
To check this we performed Monte-Carlo simulations of this model
and results for the remanent magnetization $C_T(t,0)$ at various 
temperatures are shown in figure 1. 
One observes the plateau at the value $0.475$ extending to larger 
and larger times for decreasing temperatures. For low temperatures
the final decay of $C_T(t,0)$ seems to be algebraic (before it 
will cross over to an ultimately exponential decay at times
comparable to the equilibration time, cf.\ \cite{oned}). Fitting 
a straight line to this decay in a log-log plot yields an 
intersection with the (imaginary) line $C_T=0.475$, by which we define
the time scale $t_{\rm plateau}$ for the lifetime of the metastable
state. The temperature dependence of this quantity is depicted in the insert
of figure 1 and yiels, as expected $t_{\rm plateau}\propto\tau_{\rm freeze}$.

We looked also for the waiting time ($t_w$) dependence of $C(t,t_w)$
and found that the zero-temperature-predictions (\ref{zero}) are
indeed also fulfilled for finite temperatures as long as $t<\tau_{\rm
freeze}$.  Note that obviously for $t<\tau_{\rm freeze}$ scaling laws
like $C(t,t_w)\sim\tilde{c}(t/t_w)$, which apply in many aging scenarios
\cite{oned,agescale} cannot hold.  However, for $t>\tau_{\rm freeze}$
this conventional scaling is restored.  Furthermore we calculated the
remanent energy $E_T(t)$ in Monte-Carlo simulations. The result is
depicted in figure 2. As for the remanent magnetization one sees the
characteristic initial plateau. Furthermore by plotting the value of
$E_T(t)$ versus temperature $T$ for fixed time $t$ one gets a
characteristic nonmonotonic behavior.  However, for $t\to\infty$ the
location of the minimum approaches $T=0$ and one obtains a monotonic
increase with temperature as expected for an
equilibrium-thermodynamical internal energy.

Inspecting again configuration (\ref{config}) and the following
analysis one might expect that once the observation time reaches the
time scale $\tau_{\rm freeze}$ the domain wall will perform a random
walk. In an arbitrary metastable configuration domain walls will
randomly diffuse on a characteristic time scale $\tau_{\rm freeze}$
and annihilate when two of them meet.  This scenario will result in a
$\sqrt{t}$ growth of the domain size.  To check this, we measured the
average domain size in Monte-Carlo simulations. We define size $l$ of
a domain to be equal to the number of spin pairs between to succeding
$\tau$-variables that have the value $-1$. Then, at a time $t$, we
count the number $n_l(t)$ of domains of size $l$. This defines the
probability $P_T(l,t)=ln_l(t)/L$ for a spin pair to be contained in a
segment of length $l$.

In figure 3 we show the result of the average domain size at time $t$
after the quench $d(t)=\sum_l l P_T(l,t)$ in a log-log plot. Again
one recognizes the frozen regime from $d(t)$ being constant for
$t\ll\tau_{\rm freeze}$. For larger times an intermediate growth
regime follows and we inserted a graph of $d(t)\propto{\sqrt t}$ for
comparision. One concludes that the above mentioned picture of domain
wall diffusion and annihilation is indeed to be applicable here. As
soon as $t$ reaches the order of the equilibration time $\tau_{\rm
eq}$, the domains stop to grow and $d(t)$ saturates at a value
proportional to the equilibrium correlation length $\xi_{\rm eq}(T)$,
which can be calculated analytically \cite{big}
\begin{equation}
\xi_{\rm eq}(T)=\frac{3}{2}\vert\log\,\tanh\,(J/T)\vert^{-1}
\propto \exp(J/T)\quad{\rm for}\quad T\ll J\;.
\label{corr}
\end{equation}
Apart from the prefactor $3/2$ (in general $p/2$)
the result (\ref{corr}) is identical to the case $p=2$ \cite{glauber}.
This is a general feature of the model (\ref{hamil}): although the dynamics
shows drastic differences between $p$ even and odd it turns out that
the equilibrium behavior of static quantities is very similar.

At zero temperature one again can calculate the average domain size exactly,
and beyond that also the whole probability distribution $P_{T=0}(l,t)$
for a site being contained within a domain of size $l$ at time $t$. Here
we only give the result, details of the calculation will be published
elsewhere \cite{big}:
\begin{equation}
P_{T=0}(l,t=1)=l(l-1)\,\left(\frac{1}{2}\right)^{l+2}
\quad{\rm for}\quad l\ge2
\end{equation}
and
\begin{equation}
\begin{array}{lcl}
P_{T=0}(l=2,t\ge2)&=&0\\
P_{T=0}(l=3,t\ge2)&=&\frac{9}{64}\\
P_{T=0}(l\ge4,t\ge2)&=&l\left(3l-\frac{26}{5}\right)
\left(\frac{1}{2}\right)^{l+3}
\end{array}\;.
\end{equation}
Remember that after two timesteps the system is frozen at zero temperature.
The average domain size $d_{T=0}(t)$ at zero temperature is then given by:
\begin{equation}
d_{T=0}(t=1)=5\;,\quad d_{T=0}(t\ge2)=\frac{231}{40}=5.775\;.
\end{equation}
These analytical results are compared with data obtained from
Monte-Carlo simulations in figure 4a. The agreement is excellent even
at finite temperatures for $t\ll\tau_{\rm freeze}$. In figure 4b we
show also results for $P_T(l,t)$ at higher temperatures for $t=1,2$,
an intermediate time (in the growth regime) and a time larger than the
equilibration time --- thus reflecting the equilibrium distribution.

To conclude we have presented and analyzed a simple
one-dimensional model whose non-equilibrium dynamics seems
to share many features with a glass transition. One of
them is for instance the complete freezing of the system 
in an "amorphous" state for a macroscopic time when cooled 
rapidly to low temperatures. Another is that this complex
dynamics is achieved without putting in any disorder by hand.
Of course, due to its one-dimensionality, it does not have a 
phase transition and also no particular temperature can be 
identified with a glass transition (leaving aside the question
whether the latter is a true equilibrium phase transition or 
of purely dynamical origin).
However, having demonstrated that even very simple models
yield a very rich dynamical behavior, gives us some confidence that in
higher dimensional models one might find indeed a 
candidate that shares more or even {\it all} features with a 
glass transition (as presently discussed in the
context of geometrically frustrated models \cite{chandra}).

As we have shown many new features of our model arise from the
presence of mutli-spin interactions. Thus it seems worthwhile to have
a closer look to such models in two or three dimensions, as already
discussed in \cite{riegerp}.  To support this view let us mention that
it has been pointed out several years ago \cite{kirk} that mean-field
models with $p$-spin interactions (see also \cite{pspin}) show a
dynamical behavior that is identical to that found in mode-coupling
theories of the structural glass transition.  Moreover, very recent
work on self induced disorder in models with long range interactions
\cite{self} heavily rely on multi-spin interactions, too.

This work was performed within the SFB 341
K\"oln--Aachen--J\"ulich.

\begin{figure}
\caption{The remanent magnetization for various temperatures calculated
via Monte-Carlo simulation of a system with $10^6$ spins.  From left
to right: $T=0.14$, $0.17$, $0.20$, $0.23$, $0.26$ $0.29$, $0.32$ and
$0.35$.  The insert shows the temperature dependence of $t_{\rm
plateau}$ defined in the text, the straight line is the predicted
dependency $\tau_{\rm freeze}\sim\exp(2J/T)$ ($J=1$).}
\label{fig1}
\end{figure}

\begin{figure}
\caption{The energy $E_T(t)$ for various temperatures,
>From left to right: $0.17$, $0.20$, $0.23$, $0.26$ $0.29$, $0.32$,
$0.35$, $0.4$, $0.5$, $0.6$ and $0.7$.  The insert shows the
temperature dependence of $E_T(100)$ and $E_T(500)$.}
\label{fig2}
\end{figure}

\begin{figure}
\caption{Average domain size in dependence of the waiting time $t$
in a log-log plot. The intermediate growth (between melting of the
frozen domains and final saturation by equilibration) can be fitted 
nicely to $d(t)\sim t^{1/2}$.}
\label{fig3}
\end{figure}

\begin{figure}
\caption{Probability distribution $P_T(l,t)$ for domain sizes $l$ 
at time $t$ obtained from Monte Carlo simulations. Left: $T=0.17$ and
$t=1,2$, the full lines is the analytical result. Right:
$T=0.5$ and $t=1$, $t=100$ and $t=30000\sim\tau_{\rm eq}$.}
\label{fig4}
\end{figure}


\begin{references}
\bibitem{remark} 
	For simplicity we stipulate that $L$ is a multiple of $p$.
\end{references}
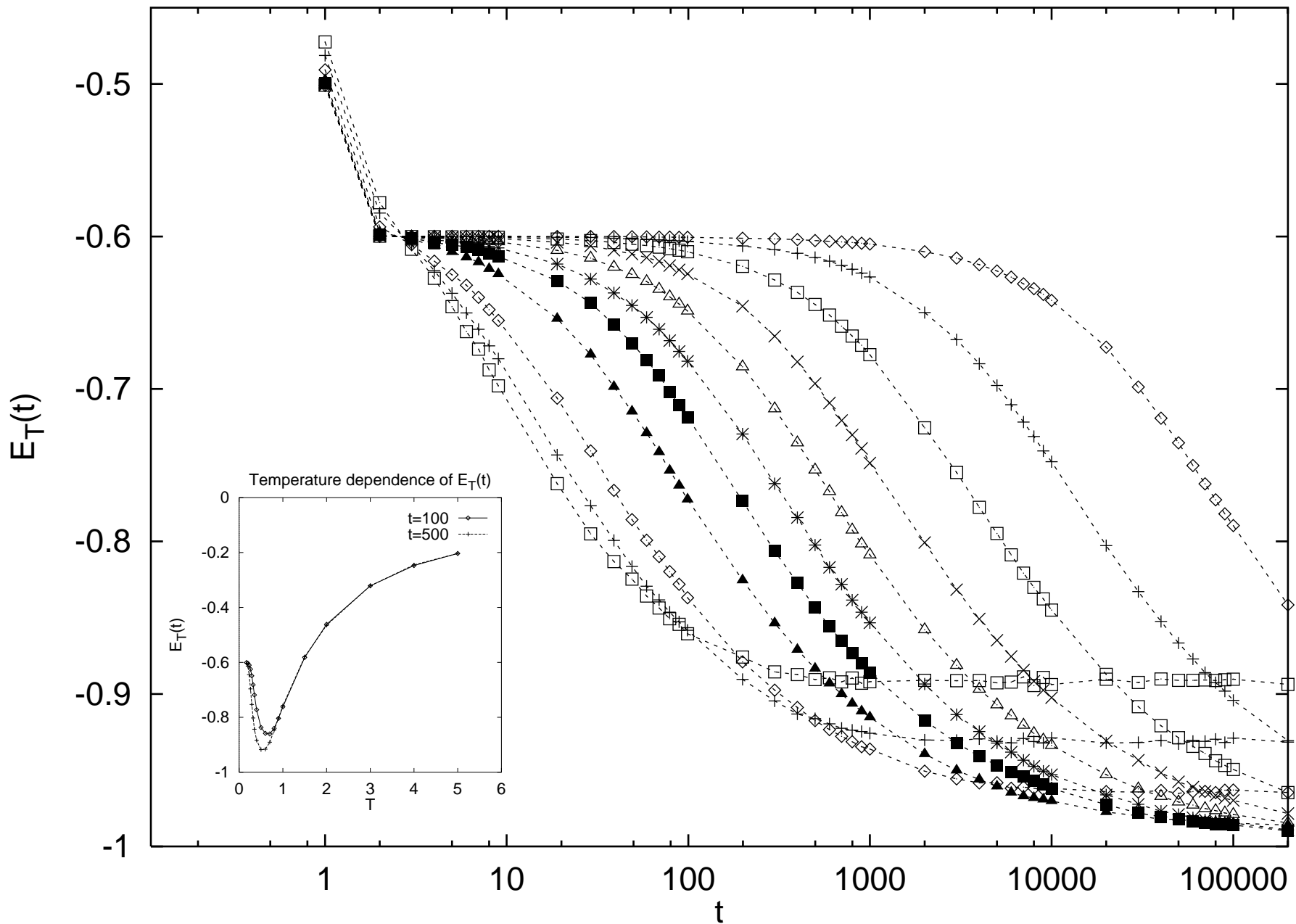
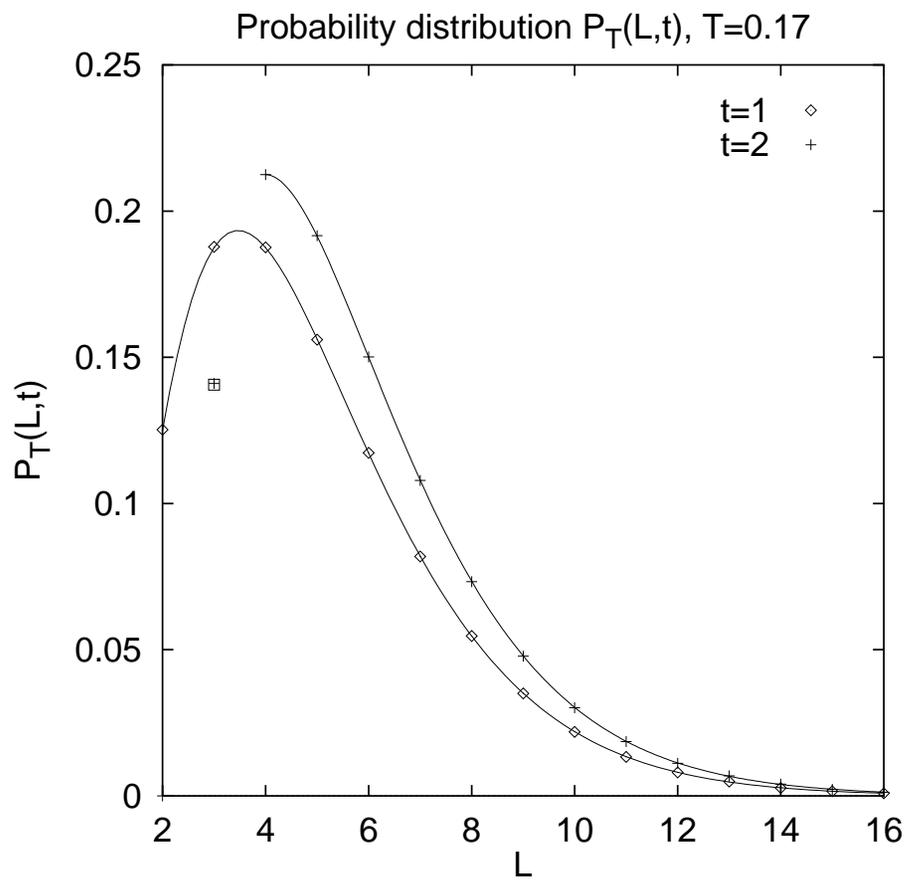
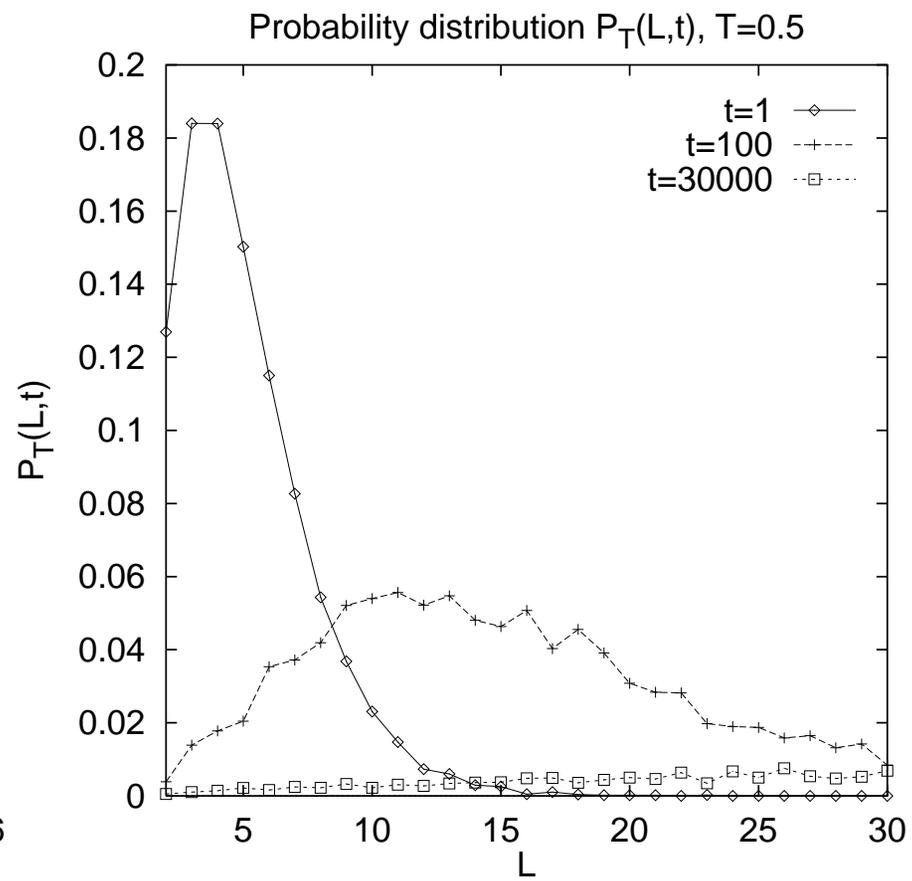
\end{document}